\documentstyle[prl,aps,twocolumn]{revtex}
\begin{document}
\draft
\title{\hfill {\small ITP Preprint Number NSF-ITP-9512}  \\
\vspace{10pt}
Spatially Ordered Fractional Quantum Hall States}
\author{Leon Balents}
\address{Institute for Theoretical Physics, University of California,
Santa Barbara, CA 93106-4030 \\ {\rm (February 2, 1995)}\\
\vspace{.2in}\begin{minipage}[t]{5.8in} \rm
 Fractional quantum Hall liquids can accomodate various degrees of
 spatial ordering.  The most likely scenarios are a Hall hexatic, Hall
 smectic, and Hall crystal, in which respectively orientational,
 one--dimensional translational, and two--dimensional translational
 symmetries are broken.  I derive the long--wavelength properties of
 these phases and the transitions between them using the Chern--Simons
 Landau--Ginzburg mapping, which relates them to spatially ordered
 superfluids.  The effects of coupling to a periodic or anisotropic
 ``substrate'' (e.g. a gate array) are also discussed.
\end{minipage}}

\maketitle

\bigskip

A general problem in condensed matter physics is the nature of quantum
solids, or more precisely, zero temperature quantum states with broken
spatial symmetries.  It has long been realized that crystalline order
could coexist with superfluidity in a bosonic
system\cite{Chester,Leggett}. Such a ``supersolid'' has, in three
dimensions, both off--diagonal long range order (ODLRO) and diagonal
long range order (DLRO) in the density matrix.  Ref.\onlinecite{TAH}\
introduced the Hall crystal, an electron crystal that nevertheless
exhibits a quantized Hall effect\cite{QHreview}.  Coexistence of
superfluidity with spatial ordering has enjoyed a recent revival of
interest\cite{FNF,PR,MSWG,BN}, in part because of the mapping of
thermally fluctuating vortex lines onto two dimensional quantum
bosons\cite{NelsonSeung}. In addition to the supersolid, under
appropriate conditions superfluidity can coexist with hexatic
(bond--orientational) order\cite{MSWG}\ and smectic (one--dimensional
translational) order\cite{BN}.

In this letter, I use the Chern--Simons Landau--Ginzburg (CSLG)
theory\cite{CSLG}\ to find fractional quantum Hall effect (FQHE) analogs
of these superfluid states.  Using the CSLG formalism, it is shown
below that the bosonic supersolid maps precisely to the Hall crystal
of Ref.\onlinecite{TAH}.  The other superfluid phases lead to new
quantum hall states: the Hall hexatic and Hall smectic.  These new
phases are probably {\sl more} likely to occur in experimental systems
than the Hall crystal.
Spatial ordering could be
encouraged by applying an external periodic potential (e.g. gate
array) to the two dimensional electron layer.  In all cases, the Hall
conductivity retains its usual quantized value.

I will consider a fermion hamiltonian of the form (in units with
$\hbar = 1$)
\begin{eqnarray}
H & = & \sum_i \left\{-{1 \over {2m}}\left( \bbox{\nabla}_i + {{ie\bbox{A}({\bf
r}_i)} \over c} \right)^2 + U_s({\bf r}_i)\right\} \nonumber
\\
& &  + {1 \over 2}\sum_{i,j} V({\bf r}_i - {\bf r}_j),
\label{fermionhamiltonian}
\end{eqnarray}
where $U_s({\bf r})$ is a periodic substrate potential, and $V({\bf r})$
is a two body interaction, which may be either unscreened Coulomb
($\sim e^2/\epsilon r$) or screened to a short range form.  The
applied magnetic field is related to the vector potential by $B
= \bbox{\nabla} \times \bbox{A}$.  The CSLG transformation defines a
bosonic wavefunction
\begin{equation}
\Psi_{\rm B}({\bf r_1,r_2,\cdots,r_N}) =
\prod_{j>k}e^{i\theta_{jk}\alpha} \Psi_{\rm F}({\bf
r_1,r_2,\cdots,r_N}),
\label{wftransform}
\end{equation}
where $\theta_{jk} = \tan^{-1}[(y_j-y_i)/(x_j-x_i)]$, and $\alpha$ is
an odd integer.  The dynamics of the transformed system is described
by the action
\begin{eqnarray}
S_{\rm B} & = & \int \! d^2{\bf r}dt \left\{ \psi_{\rm B}^\dagger \left[
i\partial_t + {1 \over {2m}} \left(\bbox{\nabla} + {{ie{\cal
A}} \over c} \right)^2 \right]\psi_{\rm B} - U_s n\right\}
\nonumber \\
& & - {1 \over 2}\int \! d^2{\bf r} d^2{\bf r'} dt \delta n({\bf r}) V({\bf r -
r'}) \delta n({\bf r'}),
\label{action}
\end{eqnarray}
where the density $n = \psi_{\rm B}^\dagger \psi_{\rm B}$, and the
$\delta n = n - \bar{n}$, with $\bar{n}$ the mean density.  The
effective vector potential ${\cal A} = A + a$, and the Chern--Simons
gauge field satisfies $\bbox{\nabla} \times a({\bf r}) = - \alpha
\phi_0 n({\bf r}) \hat{z}$, with $\phi_0 = hc/e$.  At the special
filling fractions $\bar{n}\phi_0/B = 1/\alpha = \nu$,  the mean
magnetic field vanishes, and
\begin{equation}
{\cal A}_j({\bf r}) = - {\alpha \over {2\pi}} \int \! d^2{\bf r'} \;
\delta n({\bf r'}) \epsilon_{jk} {{r_k - r_k'} \over {|{\bf r-r'}|^2}}.
\label{gaugefielddef}
\end{equation}
An important consequence  of
Eq.\ref{wftransform}\ is that the boson density $n$ is equal to
the original fermion density.  Therefore, spatial ordering of the
bosonic system implies identical ordering of the underlying electrons.

It is useful to employ a representation in terms of the
currents, in which Eq.\ref{gaugefielddef}\ becomes a consequence of
the dynamics of an appropriate Chern--Simons term.  This is achieved
by making the decomposition $\psi = \sqrt{\bar{n} + \delta
n}e^{i\phi}$, decoupling the $|\nabla\phi|^2$ term via a
Hubbard--Stratonovich field (current) $\bbox{J}$, and integrating out
the phase $\phi$.  One finds
\begin{eqnarray}
S_J  & = & \int \! d^2{\bf r}dt \bigg\{ 2\pi J \cdot c + {1
\over 2} \sum_\mu K_\mu J_\mu^2 - e J\cdot (a+A) \nonumber \\
& & + {{e^2\nu} \over {4\pi}} a\cdot (\partial\times a) \bigg\},
\label{currentaction}
\end{eqnarray}
where the 3--current $J = (n,\bbox{J})$, $a$ is the Chern--Simons
gauge field, and $c$ is a gauge field describing vortices in the
composite boson superfluid, such that the vortex current $j_v =
\partial \times c$.  For simplicity of presentation, I have made a
local approximation for the interaction $V({\bf r}) = v\delta({\bf
r})$.  A Coulomb interaction is easily accomodated by choosing
$v \propto q^{-1}$.  Within this approximation, $K_0 = -v$,
and $K_1 = K_2 = m/\bar{n}$.  In the remainder of the paper, I will
denote 3--vectors by plain characters, and ordinary spatial (two
dimensional) vectors in bold face.


Eq.\ref{currentaction}\ is ideally suited for constructing Landau
theories, because it is expressed directly in terms of physical
currents.  It must, however, be supplemented by charge conservation,
$\partial \cdot J = 0$.  Indeed, implementing this constraint via a
Lagrange multiplier field $\phi$ recovers the original phase
description.

The Hall hexatic is characterized by broken rotational (but not
translational) invariance.  For generality, I will consider a p--atic
order parameter $\psi_p({\bf r_i},\tau) = \langle \sum_{<ij>} \exp(p i
\theta_{ij})\rangle$, where $\theta_{ij}$ is the angle of the
bond between particles $i$ and $j$, and the sum is over a local set of
``neighbors.'' A more precise definition, which should be useful for
microscopic calculations, is
\begin{equation}
\psi_p({\bf r},\tau) = \left\langle \int \! d^2{\bf r'} n({\bf r},\tau)
n({\bf r'},\tau) (z-z')^p e^{-|{\bf r - r'}|^2/2} \right\rangle,
\label{hexaticdef}
\end{equation}
where $z = x + i y$.  I have been unable to find a suitable decoupling
to extract an action for $\psi_p$ directly from
Eq.\ref{action}\cite{extractnote}. Instead, in the spirit of
Landau theory, I postulate a p--atic action ($S = S_J + S_p$)
\begin{eqnarray}
S_p & = & \int \! d^2{\bf r}dt \bigg\{ {{\tilde{K}_p} \over
2}|\partial_t \psi_p|^2 - {{K_p} \over 2}|\nabla\psi_p|^2 - {{r_p}
\over 2} |\psi_p|^2
\nonumber \\
& & - {{u_p} \over 4}|\psi_p|^4 + h_s \left(\psi_p^k + {\rm
c.c.}\right) . \bigg\}
\label{hexaticaction}
\end{eqnarray}
The substrate--bond angle interaction ($\propto h_s$) is characterized
by $k$, the smallest positive integer such that $pk/m$ is also an
integer, for a substrate with $m$--fold rotational symmetry.



Because of incompressibility, additional couplings to long--wavelength
density and phase modes are irrelevant to the critical behavior (for a
true bosonic superhexatic transition, such terms are relevant for
small $k$).  When $r_p$ changes sign from positive to negative, the
system undergoes (for $u_p >0$) a second order Hall liquid to Hall
hexatic transition.  For $\lambda_s = 0$, or $k > k_c \approx 3.41$,
the transition is in the 3d XY universality class\cite{kcnote}. For
$k=2$ and $k=3$, the transitions are of the 3d Ising and 3d chiral
Potts type.  Any of these transitions can, of course, also be first
order.

For $r_p <0$, the orientational order parameter may be decomposed
$\psi_p = \psi_{0p} e^{ip\theta}$, leading via Eq.\ref{hexaticaction}\
to the low energy action
\begin{equation}
S_p = \int \! d^2{\bf r}dt \bigg\{ {\tilde{\kappa}_p \over 2} (\partial_t
\theta)^2 - {\kappa_p \over 2} |\nabla\theta|^2 + \eta_s \cos pk\theta \bigg\},
\label{bondangleaction}
\end{equation}
where $\tilde{\kappa}_p = p^2|\psi_{0p}|^2\tilde{K}_p$, $\kappa_p =
p^2|\psi_{0p}|^2 K_p$, and $\eta_s = 2 |\psi_{0p}|^k h_s$.  Note that for
$\eta_s \neq 0$, the bond--angle field $\theta$ is actually massive,
and there are no gapless modes associated with orientational
fluctuations.  The absence of these Goldstone modes is
due to the discreteness of the rotational symmetry when
substrate effects are included.

Measurable consequences of $p$--atic order arise from couplings of
$\psi_p$ to the currents.  The most relevant terms consistent with
rotational, inversion, reflection, and time--reversal symmetries are
\begin{eqnarray}
S_{J-\theta} & = & \int \! d^2{\bf r}dt \bigg\{ (g_1 + \tilde{g}_1 J_0)
{\rm Re}\left[
\psi_p^*(\partial_x +i\partial_y)^p \right]J_0 \nonumber \\
& & + g_2 {\rm Re}\left[ \psi_p^*(J_x - iJ_y)^p \right]
+ g_3 \partial_t \theta \epsilon_{\alpha\beta} \partial_\alpha
J_\beta \bigg\}.
\label{Jthetacouplings}
\end{eqnarray}
The $g_1$ and $\tilde{g}_1$ couplings reflect the anisotropy of
density fluctuations in the p--atic background, while $g_2$ allows
``easy'' propagation of currents along the p--atic axes.
The $\tilde{g}_1$ term gives rise to a $p$--fold
modulation (which vanishes as $|q|^{2+p}$ for small momentum due to
incompressibility) in the intensity of the static structure function
($\langle n({\bf q},t)n(-{\bf q},t)\rangle$) with angle.


The third ($g_3$) term in Eq.\ref{Jthetacouplings}\ is irrelevant by
power counting relative to the bond angle action,
Eq.\ref{bondangleaction}.  Nevertheless, it has an interesting
interpretation, which may give rise to physical effects away from
magic filling factors.  It exists because a non--zero angular velocity
of the bond angle field necessarily implies ``twisting'' currents,
characterized by the vorticity $\bbox{\nabla}
\times {\bf J}$.  Additional insight is gained by integrating out the
transverse part of the current.  Following Zhang\cite{CSLG}, I
introduce a gauge field $b$ satisfying $J = \partial \times b$ to
satisfy the continuity constraint, and integrate out both $a$ and
$b_0$ in the Coulomb gauge $\bbox{\nabla} \cdot {\bf b} = 0$.  The
remaining action describes density fluctuations via $\delta n =
[\bbox{\nabla} \times {\bf b}]_0$.  At long wavelengths,
the dominant remaining term is
\begin{equation}
S_{\delta n - \theta} = \int \! {{d^2{\bf q}} \over
{(2\pi)^2}}{{d\omega} \over {(2\pi)}}
{{2\pi^2\bar{n}} \over {m\nu^2q^2}} \left| \delta n - \nu n_v + {{\nu g_3}
\over {2\pi}}i\omega q^2\theta\right|^2 ,
\label{deltanthetaaction}
\end{equation}
where $n_v = j_{v,0}$ is the vortex density.
For $\theta=0$, Eq.\ref{deltanthetaaction}\ gives the usual $1/q^2$
gauge--field mediated density--density interaction term corresponding
to the incompressibility of the FQH state (where vortices carry
fractional charge $\nu$).  The cross--term coupling $\theta$ and
$\delta n - \nu\rho_v$ is completely local, and demonstrates that a
density fluctuation induces a finite local angular velocity.  This
occurs because each composite boson carries with it $1/\nu$ flux
tubes, so an increase in $n$ must be accompanied by screening
currents.  These screening currents can only be carried by the
electrons themselves, which sets the local bond angle in motion.  I
note in passing that a similar coupling between the vorticity and
bond--angular velocity also exists in a Bose superhexatic.  It would
be interesting to explore the consequences of this coupling on
the properties of vortex states in a rotating superfluid.

The greatest degree of spatial ordering possible is two--dimensional
crystallinity, and a phase in which this coexists with bosonic
off--diagonal quasi--long--range order (ODQLRO) is called a Hall
crystal.  This state was proposed in Ref.\onlinecite{TAH}\ and
analyzed using both general topological relations and a Hartree--Fock
approach on a particular model.  Here, I describe the correspondence
of the Hall crystal to a bosonic supersolid, and how the modifications
from supersolid behavior arise from Chern--Simons effects.  All the
predictions of the Chern--Simons theory are in accord with the earlier
work of Ref.\onlinecite{TAH}.

A number of authors have studied the existence of crystalline order in
a superfluid state\cite{Chester,Leggett,FNF,PR}. It is signaled by
non--zero expectation values of the Fourier components of the density,
\begin{equation}
n_{\bf G} = \sum_j \exp\left(i{\bf G}\cdot {\bf r}_j\right) ,
\label{top}
\end{equation}
for all ${\bf G}$ in the reciprocal lattice.  This is precisely the
additional order parameter of the Hall crystal given in
Ref.\onlinecite{TAH}.

To study the Hall crystal, I assume small phase fluctuations in the
translational order parameters via $n_{\bf G} = n_{\bf G 0}\exp(i{\bf
G}\cdot{\bf u})$.  The ${\bf u}$ field represents the Goldstone modes
of broken translational invariance, i.e. phonons.  For a high symmetry
(e.g. hexagonal) crystal, it takes the form
\begin{equation}
S_{u} = \int \! d^2{\bf r}dt \bigg\{ {\sigma \over 2} |\partial_t
{\bf u}|^2 -\mu u_{\alpha\beta}^2 - {\lambda \over 2}
u_{\alpha\alpha}^2 \bigg\},
\label{phononaction}
\end{equation}
where $u_{\alpha\beta} = (\partial_\alpha u_\beta + \partial_\beta
u_\alpha)/2$ is the strain tensor.  For simplicity, I have assumed there
are no couplings to the substrate lattice, i.e. that the density wave is
incommensurate.  Such terms generally create a gap for phonons.

It is important to note that the definition of translational order in
Eq.\ref{top}\ implies only short--wavelength (at wavevector ${\bf Q}$)
deformations of the density.  In particular, the usual association of
a compressional phonon with a change in the long--wavelength density
need not hold, i.e. $\delta n + n_0 u_{\alpha\alpha} \neq 0$.  This
difference, which in conventional crystals is small at finite
temperature and strictly zero at zero temperature, can be understood
as arising due to a non--zero concentration of vacancies or
interstials in the crystal.  A longitudinal phonon may be accompanied by
an increase in the local vacancy density to keep the mean density
uniform.

For the more relevant case of a {\sl weak} density wave, where
$\langle n_{\bf Q} \rangle \ll n_0$, it is more appropriate to
describe the system in terms of a compressional phonon mode and a
true long--wavelength density fluctuation (which can be decomposed
into a linear combination of the defect and lattice densities as
suggested above).  The modes are coupled via interactions of the form
\begin{equation}
S_{J-{\bf u}} = \int \! d^2{\bf r}dt \bigg\{ h_1
u_{\alpha\alpha}\delta n + {h_2 \over 2} \partial_0 {\bf
u} \cdot {\bf J} \bigg\}.
\label{Jucoupling}
\end{equation}
In a true bosonic supersolid, $h_1$ and $h_2$ would mix the
long--wavelength density fluctuations (sound waves) and lattice
compressions, giving two new linearly dispersing modes.  Because of
incompressibility, however, a lattice compression $u_{\alpha\alpha}$
does not lead to any change in the long--wavelength density $\delta n$
as $q \rightarrow 0$ in the Hall crystal (i.e. there remains a charge
gap).

Following the methods used to derive Eq.\ref{deltanthetaaction}\ and
shifting $\delta n \rightarrow \nu\delta n_v + \nu h_2
\partial_0 \theta/(2\pi)$ gives the form of the vortex--phonon interactions,
\begin{equation}
S_{n_v - {\bf u}} = \int \! d^2{\bf r}dt \bigg\{ \tilde{h}_1 n_v
u_{\alpha\alpha} + \tilde{h}_2\left[K_0 + K_1 {\partial_0^2 \over
\nabla^2}\right] n_v \partial_0
\theta \bigg\},
\label{nvucoupling}
\end{equation}
where $\theta = \epsilon_{\alpha\beta}\partial_\alpha u_\beta$ is the
local bond angle in the lattice, $\tilde{h}_1 = \nu h_1$, and
$\tilde{h}_2 = \nu^2 h_2/(2\pi)$.  The first coupling describes the
tendency of the lattice to stretch when a vortex (fractionally charged
quasiparticle) is added to the system, while the second is present due
to circulating screening currents (see the discussion after
Eq.\ref{deltanthetaaction}).  Mathematically similar terms have been
studied in chiral polymer assemblies, where the incompatibility of
such local rotation of the bond angle with crystallinity leads to very
complex spatial structures\cite{KN}. Somewhat analogous physics might
occur away from magic filling factors in a sufficiently clean sample.

As discussed in Ref.\onlinecite{BN}, anisotropy (from, e.g., an
anisotropic effective mass tensor or a periodic external gate array)
encourages the formation of a one--dimensional charge density wave
(DW).  I will denote such a state as a Hall smectic, by analogy with
one--dimensionally ordered liquid crystals\cite{PdG}. It is
intermediate between the Hall crystal and liquid (or hexatic).  I
briefly summarize its properties, which may be derived analogously to
those of the Hall crystal\cite{Bunpub}. It can arise via a second
order transition from the liquid or $p$--atic phase.  The critical
behavior (for a featureless or small--scale periodic substrate) is in
the XY universality class, governed by the single complex order
parameter $n_{\bf G}$, where ${\bf G}$ is the wavevector of the DW.
The low energy spectrum contains a single longitudinal phonon mode
with linear dispersion.  Vortices couple to this mode exactly as in
Eq.\ref{nvucoupling}, where now $\theta \propto
\epsilon_{\alpha\beta} G_\alpha \partial_\beta u$.

Much of the progress in understanding the FQHE occurred through the
analysis of trial wavefunctions.  What is an appropriate form for such
a spatially ordered Hall system?  The remarkable answer is that the
original Laughlin wavefunctions actually describe these states for
certain (large) filling factors\cite{Laughlin}.  This can be see
using the well--known ``plasma analogy''\cite{plasma,chesternote}.   The
probability density in the Laughlin state is
\begin{equation}
P_{\rm L} = |\psi_{\rm L}|^2 = e^{-U/k_{\rm B}T},
\label{Lprob}
\end{equation}
where $k_{\rm B}T = \nu$ and
\begin{equation}
U = -2q^2 \sum_{i<j} \ln |z_i - z_j| + {q \over 2} \sum_i |z_i|^2,
\label{plasmaenergy}
\end{equation}
with $q = 1/\nu$.  Evaluating a correlation function involving
densities amounts to computing a classical statistical mechanical
average in an $N$ particle system with the energy $U$.
Eq.\ref{plasmaenergy}\ implies that these particles interact like
charges in a two dimensions.  Such logarithmically interacting
particles have been well--studied in the context of two--dimensional
vortex films, from which it is known that they form a
quasi--long--range ordered solid at low temperatures.  Laughlin's
wavefunction therefore describes a state of broken translational
symmetry for $\nu < \nu_m$, where $1/109 \lesssim \nu_m \lesssim
1/59$\cite{Calliol,DSFfilm,Bunpub}. Provided melting occurs via a continuous
or sufficiently weakly first order transition, the system will sustain
a hexatic phase for $\nu \gtrsim \nu_m$\cite{hexaticnote}.
Since Laughlin's state quite generally exhibits the FQHE, his
wavefunction in this regime must represent a Hall crystal and Hall
hexatic.  Of course, current estimates which place the critical
filling fraction for Wigner crystallization at $\nu \approx
1/7$ demonstrate that Laughlin's wavefunction is no longer a good
approximation in this regime\cite{LGWigner}.  Other wavefunctions
may, however, display spatial order at larger fillings.

What are the chances of seeing these exotic phases experimentally?
The absence of demonstrated sub--monolayer spatial ordering in $^4$He
films suggests it may be difficult\cite{Mochelnote}. The helium
experiments, are, however, complicated by problems of phase separation
and relatively high particle density.  The long range, purely
repulsive electron--electron interaction should prohibit phase
separation.  The low carrier density allows, in principle, the
fabrication of a periodic gate array to generate anisotropy and
encourage ordering at particular wavevectors.  Such a periodic
potential has been argued (in ways which should carry over to
the FQHE case) both theoretically\cite{BN}\ and numerically\cite{ZSS}\
to encourage supersolid order in helium films.  Observation of any
spatially ordered state is limited by the effects of randomly
distributed donor impurities.  General arguments show that such
disorder always destabilizes incommensurate spatial
ordering\cite{FLee}, leading to finite translational and orientational
correlation lengths, as for the Wigner crystal\cite{Fertig}.  A
commensurate periodic substrate potential, however, allows true long
range spatial order even in the presence of a weak impurity potential.

A final, intriguing possibility is suggested by comparison with helium
films with many monolayers of coverage.  Such systems exhibit
supersolid order by having a number of ``inert'' crystalline layers
underneath a superfluid film.  Although it may appear an uninteresting
example, exchange of atoms between the inert and superfluid layers
implies that the state is still non--trivial.  A somewhat analogous
situation might be achieved in a double layer quantum Hall
heterostructure.  By making the two layers inequivalent (either
through fabrication or a bias voltage), a Wigner crystal could be
formed in one layer while the other contains a Laughlin liquid.  In
the absence of tunneling and interactions, this situation is
technically a ``partial'' Hall crystal, as defined in
Ref.\onlinecite{TAH}.  It would be interesting to explore the
consequence of couplings between the two layers on such a state.

It is a pleasure to acknowledge discussions with Matthew Fisher,
Steve Simon, and Andrew Tikofsky.  This research was supported by the
National Science Foundation under Grant No. PHY89--04035 at the ITP.


\begin{references}

\bibitem{Chester} G. V. Chester, Phys. Rev. A {\bf 2}, 256 (1970).

\bibitem{Leggett} A. J. Leggett, Phys. Rev. Lett. {\bf 25}, 1543 (1970).

\bibitem{TAH} Z. Tesanovic, F. Axel, and B. I. Halperin, Phys. Rev. B
{\bf 39}, 8525 (1989).

\bibitem{QHreview} A. H. MacDonald, unpublished.

\bibitem{FNF} E. Frey, D. R. Nelson and D. S. Fisher, Phys. Rev. B{\bf
49}, 9723 (1994).

\bibitem{PR} Y. Pomeau and S. Rica, Phys. Rev. Lett. {\bf 72}, 2426
(1994).

\bibitem{MSWG} K. Mullen, H. T. C. Stoof, M. Wallin, and S. M. Girvin,
Phys. Rev. Lett. {\bf 72}, 4013 (1994).

\bibitem{BN} L. Balents and D. R. Nelson, Phys. Rev. Lett.  {\bf
73}, 2618 (1994); see also unpublished.

\bibitem{NelsonSeung} D. R. Nelson, Phys. Rev. Lett. {\bf 60}, 1415
(1988); D. R. Nelson and S. Seung, Phys. Rev. B{\bf 39}, 9153 (1989).

\bibitem{CSLG} For a review, see S. C. Zhang, Int. J. Mod. Phys. B
{\bf 6}, 25 (1992).

\bibitem{extractnote} To my knowledge, no such treatment exists even
for {\sl classical} hexatics.

\bibitem{kcnote} See, e.g. Ref.\onlinecite{BN}.

\bibitem{KN} R. D. Kamien and D. R. Nelson, unpublished.

\bibitem{PdG} P. G. de Gennes and J. Prost, {\em The Physics of
Liquid Crystals,} (Oxford University Press, New York, 1993).

\bibitem{Bunpub} L. Balents, unpublished.

\bibitem{Laughlin} R. B. Laughlin, Phys. Rev. Lett. {\bf 51}, 605 (1983).

\bibitem{plasma} See, e.g. Ref.\onlinecite{QHreview}.

\bibitem{chesternote} Chester's demonstration of translational order
in his wavefunctions employed an analogous argument!

\bibitem{Calliol} J. M. Calliol et. al., J. Stat. Phys. {\bf 28}, 325.

\bibitem{DSFfilm} D. S. Fisher, Phys. Rev. B{\bf 22}, 1190 (1980).

\bibitem{hexaticnote} I am not aware of any quantitative estimates
of the hexatic--liquid critical temperature in the vortex system.

\bibitem{LGWigner} P. K. Lam and S. M. Girvin, Phys. Rev. B {\bf 30},
473 (1984).

\bibitem{Mochelnote} The best candidate experiment is described in
Ref.\onlinecite{Mochel}, which may have observed a
superhexatic\cite{MSWG}\ or supersolid\cite{Bunpub}.

\bibitem{ZSS} R. T. Scalettar, G. G. Batrouni, A. P. Kampf, and
G. T. Zimanyi, preprint (1994).

\bibitem{FLee} H. Fukuyama and P. A. Lee, Phys. Rev. B {\bf 17}, 535
(1978).

\bibitem{Fertig} See, e.g., M.--C. Cha and H. A. Fertig, Phys. Rev. B
{\bf 50}, 14368 (1994).

\bibitem{Mochel} M. T. Chen, J. Roesler, and J. M. Mochel, J. Low
Temp. Phys. {\bf 89}, 125 (1992); J. M. Mochel and M. T. Chen, Physica
(Amsterdam) {\bf 197B}, 278 (1994).



\end{references}
 \end{document}